\magnification=\magstep1
\font\small=cmr5
\hsize=15.6 truecm
\vsize=21.5 truecm
\baselineskip=14 pt
\tolerance=200

\def\parag{\hfil\break}
\newcount\ch 
\newcount\eq 
\newcount\foo 
\newcount\ref 

\def\chapter#1{\parag {\bf\the\ch.\enskip#1}\global\advance\ch by 1
}

\def\equation{
\eqno(\the\eq)\global\advance\eq by 1
}

\def\reference{
\parag [\number\ref]\ \advance\ref by 1
}
\def\foot#1{
\footnote{($^{\the\foo}$)}{#1}\advance\foo by 1
} 

\ch = 1 
\eq = 1 
\foo = 1 
\ref = 1 
\def\d{\partial}
\def\bR{{\rm{\bf R}}}
\def\bZ{{\rm{\bf Z}}}
\def\dslash{{\partial\mkern-2mu\llap{{
\raise+0.5pt\hbox{\big/}}}\mkern+2mu}\ }
\null \vfill
{\sevenrm\hbox to 15truecm{ISSN 0133--462X\hfil ITP
 Budapest Report No. 519}
\hbox to 15truecm{hep-th/9609198 \hfil September 1996}}
 
\centerline{$\phantom{\pi}$}
 
\vskip 1.5truecm
 
\centerline{\bf On quantum equivalence of dual sigma models:}
\centerline{\bf $SL(3)$ examples}

\vskip 2 truecm

\centerline{Z. Horv\'ath, R.L. Karp and L. Palla}
\centerline{Institute for Theoretical Physics}
\centerline{Roland E\"otv\"os University}
\centerline{H-1088 Budapest, Puskin u. 5-7, Hungary}
\vskip 4.5truecm
 
\centerline{{\bf Abstract}}

The equivalence of several $SL(3)$ sigma models and their
special Abelian duals is investigated 
in the two loop order of perturbation theory. The investigation is 
based on extracting and 
comparing various $\beta$ functions of the original and dual models. 
The role of the discrete 
global symmetries is emphasized.    
\vfill
\eject

\chapter{Introduction}

Various  target space  duality transformations [1-3] (\lq T duality')  
connecting two seemingly
different sigma-models
or string-backgrounds are playing an increasingly important role        
nowadays, since they are assumed to lead to alternative descriptions of the
same physical system. 
These transformations are the generalizations
of the $R\rightarrow 1/R$ duality in toroidal compactification of string 
theory, and are usually described in the sigma model formulation 
of the corresponding conformal field theory.

Using the sigma model formualtion 
it has been shown recently, both in the Abelian [4], and in the non Abelian 
case [5],
that the duality transformation rules can be recovered in an elegant way by
performing a canonical transformation. This clearly shows that the models
related by these transformations are equivalent {\sl classically}. In the 
quantum theory, the usual way to show that the models related by the duality 
transformations are equivalent,  
-- in spite of the generally non linear change of variables 
they involve --,  is by making some formal manipulations in the 
functional integral [1,6], ignoring the need for regularization. 
While this may be sufficient for conformal  invariant string backgrounds when
no perturbative quantum corrections are expected, 
 we feel, that from a pure $2$d field theory point of wiev,  
the question of quantum equivalence between  sigma
models  related by duality transformations deserves further study. 

 We  initiated such a study in ref.[7], where   
 the various sigma models were treated as "ordinary" 
(i.e. not necessary conformally 
invariant) two dimensional quantum field theories in the framework of 
perturbation theory. Working in a field theoretic rather than string
theoretic framework i.e. working with a flat, non dynamical $2$ space 
and ignoring the dilaton,  it was shown on a 
number of examples that the `naive' (tree level) T-duality
transformations in 2d $\sigma$-models cannot be
exact symmetries of the quantum theory. The `naive' Abelian duality
transformations 
are correct to one loop in perturbation theory
\foot{see also ref.[8]}, they break down in general,
however, at the two loop order. We reached these conclusions by analyzing 
and comparing various $\beta$ functions in the original and dual theories.

All the models investigated in ref.[7], -- the deformed principal sigma model 
and its
various duals --, were based on the group $SU(2)$. 
The aim of this paper is to 
check some of the ideas inferred from the study of these models in a slightly 
more complicated situation, i.e. when $SU(2)$ is replaced by $SL(3)$. 
In the very simple case of the
Abelian duality transformations, when the distinguished component of the 
original metric, $g_{00}$, (see eq.(3) below) is {\sl constant}, 
the standard derivation using the functional integral 
 amounts to just a standard gaussian integration, thus 
no problems are expected with the quantum equivalence of the dual theory.
In ref.[7] it was found that in this case the dual model
is indeed equivalent to two loops to the original one, however,
there is a nontrivial
change of scheme involved when insisting on dimensional regularization. We 
investigate below whether the same is true for the various deformed 
$SL(3)$ principal sigma models, when several, formally different 
dual models can 
be constructed with constant $g_{00}$. In this study we also clarify the 
role of the global symmetries -- in particular the discrete ones.    

The paper is organized as follows: in chapter 2. we give a brief review of
the duality transformations and the coupling constant renormalization 
procedure we use in the rest of the paper. In the first section of chapter 3.
the Lagrangians of the various deformed $SL(3)$ principal models, admitting
various degrees of discrete symmetries, are constructed, and in the second
section of this chapter we derive the $\beta$ functions of the most symmetric 
of them. In chapter 4.  we study the renormalization
of the two simplest duals of this model and show that the two loop counterterms
contain new terms that cannot be explained by field redefinition thus naively
these dual models are not renormalizable. However in the first section of 
chapter 5. we exhibit that working in an appropriately enlarged parameter space 
restores the renormalizability of the dual models and in the second section
we indicate -- using a generalized \lq fixed point' analysis -- that even the
physical $\beta$ functions of the original and dual models coincide, thereby 
confirming their equivalence. We summarize and make our conclusions in 
chapter 6. 
The somewhat complicated expressions of the various two loop
beta functions for the various models in the enlarged parameter space are 
collected in Appendix A, while the derivation of the two loop renorm 
invariants is sketched in Appendix B.
   
\chapter{Abelian T duality and
coupling constant renormalization}
\bigskip

We  start with a brief summary
of the Abelian T-duality [1,2,4]. To this end
consider the $\sigma$-model action:
$$
\eqalign{
S={1\over4\pi\alpha'}\int d^2z
\biggl[&\sqrt{h}h^{\mu\nu}\left(g_{00}\d_\mu\theta \d_\nu\theta
+2g_{0\alpha}\d_\mu\theta\d_\nu \xi^{\alpha} 
+g_{\alpha\beta}\d_\mu
\xi^{\alpha}\d_\nu \xi^{\beta}\right)\cr
&+\epsilon^{\mu\nu}(2b_{0\alpha}\d_\mu\theta \partial_\nu \xi^{\alpha}
+b_{\alpha\beta}\partial_\mu
\xi^{\alpha}\partial_\nu \xi^{\beta})\biggr]
}
\equation
$$
where $g_{ij}$ is the target space metric, $b_{ij}$ the
torsion potential, and the target space indices are decomposed 
as $i=(0,\alpha)$
corresponding to the coordinate decomposition $\xi^i=(\theta, \xi^\alpha)$.
The target space metric and $b_{ij}$ are assumed to possess a Killing
vector and are now written in the adapted coordinate system,
i.e.\ they are independent of the coordinate $\theta$.
$h_{\mu\nu}$ is the world sheet metric and
$\alpha^{'}$ the inverse of the string tension.
The dilaton field  is missing in eq.(1) as usual 
for non conformal sigma models: in this paper we are concerned mainly with
asymptotically free models that are believed to 
generate non zero masses by dimensional
transmutation when quantized as ordinary quantum field  theories. In the
same spirit 
 the world sheet metric, $h_{\mu\nu}$, is taken to be
flat in what follows.  (e.g.~that of a torus, to regulate the infrared 
divergences).
Writing the integrand in eq.(1) as 
$$
{\cal L}= g_{00} (\dot{\theta}^2-\theta^\prime\,^2)+(\dot{\theta}+
\theta^\prime)J_-+(\dot{\theta}-\theta^\prime)J_++V,
\equation
$$
with 
$$\eqalign{
&J_-= (g_{0i}+b_{0i})\partial_-\xi^i, \qquad
J_+= (g_{0i}-b_{0i})\partial_+\xi^i ,\cr
&V= (g_{ij}+b_{ij})\partial_+\xi^i\partial_-\xi^j,\qquad z^\pm=z^0\pm z^1,}
$$
we obtain the abelian dual with respect to the $\theta\rightarrow 
\theta +\alpha$ isometry by
performing the canonical transformation [4]:
$$
p_\theta=-{\tilde \theta}^\prime, \qquad
p_{{\tilde \theta}}=-\theta^\prime .
$$
This transformation is  
mapping the `original' $\sigma$-model with action,
$S[g\,,b]$, given in eq.~(1) to its dual, $S[\tilde g\,,\tilde b]$, 
where
$$\eqalign{
&{\tilde g}_{00}={1\over g_{00}}\qquad
{\tilde g}_{0\alpha}={b_{0\alpha} \over g_{00}},
\qquad
{\tilde b}_{0\alpha}={g_{0\alpha} \over g_{00}} \cr
&{\tilde g}_{\alpha\beta}=g_{\alpha\beta} -
{g_{0\alpha}g_{0\beta} - b_{0\alpha} b_{0\beta}\over g_{00}}\cr
&{\tilde b}_{\alpha\beta}=b_{\alpha\beta}-{g_{0\alpha}b_{0\beta}
         -g_{0\beta}b_{0\alpha}\over g_{00}}.
\cr}
\equation
$$
The seemingly different models, described by 
$S[g\,,b]$ and $S[\tilde g\,,\tilde b]$,  are classically equivalent as 
they are related by a canonical transformation. 
In principle, a possible way to
investigate the equivalence of their quantized versions is by computing
and comparing some \lq physical' quantities up to an as high order 
of perturbative expansion as possible. (In practice, for sigma models 
with torsion,  this is the two loop order).  The perturbative 
determination of quantities like the free energy density or some scattering
cross sections is greatly simplified if the model admits a sufficient degree
of symmetry. Therefore we choose the original models 
to have enough symmetry to guarantee that they 
admit a parametrization, where their complete renormalization amounts to a
multiplicative renormalization of the coupling constant(s) and the 
parameter(s); thus enabling the derivation of the corresponding $\beta$ 
functions \foot{The difference between a coupling constant and
a parameter is that the parameter is not necessarily small, thus we do not 
expand anything in the parameter, while in the 
coupling constant we assume a perturbative expansion.}. The derivation of 
these $\beta$ functions is the first step in the program of computing the
aforementioned \lq physical' quantities. Thus, as a first step, we inquire   
whether the $\beta$ functions, extracted from the dual 
models, are the same as the original ones. 

Our general strategy to carry out the renormalization of the
`original' and of the `dual' models
and to obtain the corresponding $\beta$ functions is described in some details
in ref.[7], thus here we just summarize what we need in the sequel. The 
procedure
is based on the one resp.~two
loop counterterms for the general $\sigma$-models (either with or without
the torsion term) computed by several authors [9-11]. 
These counterterms  were
derived by the background field method in the dimensional
regularization scheme. 
Writing the general $\sigma$-model Lagrangian in the form
$$
{\cal L}={1\over2\lambda}\bigl( g_{ij}(\xi)
+b_{ij}(\xi)\bigr)\Xi^{ij}
={1\over\lambda}\tilde{\cal L} ,\qquad 
\Xi^{ij}=(\d_\mu\xi^i\d^\mu\xi^j
+\epsilon_{\mu\nu}\d^\mu\xi^i\d^\nu\xi^j)
\equation
$$
and expressing the loop expansion parameter, $\alpha^\prime$, in terms of
the coupling constant, $\lambda$ as $\alpha^\prime=\lambda/(2\pi)$, the 
simple pole parts of the one
($i=1$) and two ($i=2$) loop countertems, ${\cal L}_i$, apart from the 
$\mu^{-\epsilon}$ factor, 
are given as:
$$
\mu^{\epsilon}{\cal L}_1={\alpha^\prime\over2\epsilon\lambda}
\hat{R}_{ij}\Xi^{ij}
={1\over\pi\epsilon}\Sigma_1 ,
\equation
$$
and
$$
\mu^{\epsilon}{\cal L}_2={1\over2\epsilon}
({\alpha^\prime\over2})^2{1\over2\lambda}
Y^{lmk}_{\ \ \ j}\hat{R}_{iklm}\Xi^{ij} 
={\lambda\over8\pi^2\epsilon}\Sigma_2, 
\equation
$$
where
$$
\eqalign{
Y_{lmkj}&=-2\hat{R}_{lmkj}+3\hat{R}_{[klm]j}+2(H^2)_{kl}g_{mj}-2(H^2)_{km}
g_{lj},\cr
(H^2)_{ij}&=H_{ikl}H_j^{kl},\quad 2H_{ijk}=\partial_ib_{jk}+{\rm cyclic},\cr}
\equation
$$
In these equations $\hat{R}_{iklm}$ resp. $\hat{R}_{ij}$ denote  
the generalized Riemann resp. Ricci tensors of 
 the generalized connection, $G^i_{jk}$, containing also the torsion term 
in addition to the Christoffel
symbols $\Gamma^i_{jk}$, of the metric $g_{ij}$:
 $G^i_{jk}=\Gamma^i_{jk}+H^i_{jk}$. 

 If the metric, $g_{ij}$, and the torsion 
potential, $b_{ij}$, depend also on a parameter, $x$;
$g_{ij}=g_{ij}(\xi ,x)$, $b_{ij}=b_{ij}(\xi ,x)$ then we convert the 
previous counterterms into coupling constant and parameter renormalization
by assuming that in the one ($i=1$) and two ($i=2$) loop orders
their bare and renormalized values are related as
$$\eqalign{
\lambda_0&=\mu^\epsilon\lambda\Bigl( 1+{\zeta_1(x)\lambda\over\pi\epsilon}
+{\zeta_2(x)\lambda^2\over8\pi^2\epsilon}+...\Bigr)=\mu^\epsilon\lambda
Z_{\lambda}(x,\lambda)=\mu^\epsilon\lambda\Bigl(
1+{y_\lambda (\lambda ,x)\over\epsilon}+...\Bigr)
,\cr   
x_0&= x+{x_1(x)\lambda\over\pi\epsilon}
+{x_2(x)\lambda^2\over8\pi^2\epsilon}+...=xZ_x(x,\lambda)=x\Bigl(
1+{y_x(\lambda ,x)\over\epsilon}+...\Bigr)
,\cr}
\equation
$$
(where the ellipses stand for both the higher loop contributions and for the
higher order pole terms).  
We determine the unknown  functions, $\zeta_i(x)$ and $x_i(x)$, ($i=1,2$), 
from the equations:
$$
-\zeta_i(x)\tilde{\cal L}+{\d \tilde{\cal L}\over\d x}x_i(x)+
{\delta \tilde{\cal L}\over\delta\xi^k}\xi^k_i(\xi ,x)=\Sigma_i,\qquad i=1,2.
\equation
$$
These equations express  the finiteness of the generalized quantum
effective action, $\Gamma(\xi)$, as defined in ref.s[12], 
up to the corresponding  orders in perturbation theory.
In eq.(9) $\xi^k_i(\xi ,x)$ may depend in an arbitrary way 
on the parameter, $x$,
and on the fields, $\xi^j$, the only requirement is that $\xi^k_i(\xi ,x)$ may 
contain no derivatives of $\xi^j$. Eq.(9) admits a simple
interpretation: it suggests that the general counterterms of the sigma models
may be accounted for by the coupling constant and parameter renormalization
if the latter ones are accompanied by a (in general non-linear) redefinition
of the fields $\xi^j$:
$$  
\xi^j_0=\xi^j+{\xi^j_1(\xi^k,x)\lambda\over\pi\epsilon}
+{\xi^j_2(\xi^k,x)\lambda^2\over8\pi^2\epsilon}+...,
\equation
$$
in such a way that eq.(9) holds \foot{Note that here $\xi^j$ is not assumed
to solve any equation of motion so it is not related directly to the \lq 
classical' field of ref.s[12]}. In the special case when $\xi_i^k$ depends 
linearly on the $\xi $ fields, $\xi_i^k(\xi ,x)=\xi^k y_i^k(x)$, eq.(10) 
simplifies to an ordinary multiplicative wave function renormalization.  
We emphasize that it is not guaranteed a priori that eq.(9) may be solved
for $\zeta_i(x)$, $x_i(x)$ and $\xi^k_i(\xi ,x)$. If
$\Sigma_i$  happen to have a form that makes this impossible then this implies,
that the renormalization of the model drives it - in the infinite dimensional
space of all metrics and torsion potentials - out of the lower dimensional
subspace characterized by the coupling constant(s) and the parameter(s); i.e.
implies that the model is not renormalizable in the ordinary, field theoretical
sense. On the other hand, if eq.(9) admits a solution, then, writing 
$Z_\lambda=1+{y_\lambda (\lambda ,x)\over\epsilon}+...$ and 
$Z_x=1+{y_x(\lambda ,x)\over\epsilon}+...$, the $\beta$ functions of $\lambda$ 
and $x$, defined in the standard way, are readily obtained:
$$
\beta_\lambda=\lambda^2{\d y_\lambda\over\d\lambda},\qquad 
\beta_x=x\lambda {\d y_x\over\d\lambda}.
\equation
$$
This framework was used in [7] in several examples of $SU(2)$-based sigma 
models to investigate whether
the same $\beta$ functions can be extracted from the original and the dual
models.

\bigskip
\chapter{\lq Deformed' $SL(3)$ principal sigma models}
\bigskip

\underbar{3.1 General considerations}

Below we consider various versions of the deformed $SL(3)$ principal sigma
model and its abelian duals, 
although models built from the elements of $SU(3)$ were  
the most natural generalizations of the examples investigated in ref.[7]. The 
reason for this is twofold: on the one hand working with the non compact 
version makes it possible to use the Gauss decomposition to parametrize the 
group elements, which yields, eventually, an action {\sl polynomial} (rather
than trigonometric) in the fields; while, on the other, we know [13] that
the compact and non compact versions are on the same footing  
as far as the beta functions are concerned. Therefore we parametrize 
$G\in SL(3)$ by a lower triangular, an upper triangular and a diagonal 
matrix as 
$$
G=G_UG_LG_D
\equation
$$ 
where
$$
G_U=\pmatrix{1&{\rm F_6}(x,t)&{\rm F_8}(x,t)\cr
             0&1&{\rm F_7}(x,t)\cr
             0&0&1},
\qquad
G_L=\pmatrix{1&0&0\cr
             {\rm F_3}(x,t)&1&0\cr
         {\rm F_5}(x,t)&{\rm F_4}(x,t)&1},
\equation
$$
$$
G_D=\pmatrix{e^{{\rm F_1}(x,t)+{{\rm F_2}(x,t)\over
\sqrt {3}}}&0&0\cr
0&e^{-{\rm F_1}(x,t)+{{\rm F_2}(x,t)\over
\sqrt {3}}}&0\cr
0&0&e^{-{2{\rm F_2}(x,t)\over
\sqrt {3}}}},
\equation
$$
and use ${\rm F}_i(x,t)$ ($i=1,...,8$) as the fields $\xi_i$. 

To construct the \lq original' model we follow the philosophy of ref.[7] 
in choosing it as symmetric as possible, since in this way we can keep the
number of parameters that require renormalization small. Therefore we build
the models we start with from the various components of the current, 
$J_\mu=G^{-1}\d_\mu G=J_\mu^a\lambda^a$:
$$
J_\mu=\pmatrix{ J_\mu^3+{1\over\sqrt{3}}J_\mu^8 &J_\mu^1&J_\mu^4 \cr
J_\mu^2&-J_\mu^3+{1\over\sqrt{3}}J_\mu^8 &J_\mu^6 \cr
J_\mu^5&J_\mu^7&-{2\over\sqrt{3}}J_\mu^8},
\equation
$$
since this way invariance under the \lq left' $SL(3)$ transformations,
 generated by 
constant $G_0$ elements of $SL(3)$, 
multiplying $G$, eq.(12), from the left, $G\rightarrow
G_0G$, is automatically guaranteed. The Lagrangians of the 
various \lq deformed' principal 
models are obtained by adding various quadratic expressions,  
$J_\mu^aJ^{\mu b}$, to the Lagrangian of the principal model:
$$
{\cal L}_{\rm pr}=-{1\over2\lambda}{\rm Tr}J_\mu J^\mu ,
\equation
$$
so as to break its invariance under the \lq right' $SL(3)$ transformations
($G\rightarrow GG_0$) to an appropriate subgroup. We certainly want this 
unbroken right invariance group to contain $\bR\times\bR=\bR^2$, formed by the 
${\rm F}_1\rightarrow {\rm F}_1+\alpha$ and 
${\rm F}_2\rightarrow {\rm F}_2+\beta$ translations of 
${\rm F}_1$ and ${\rm F}_2$, since we need these invariances to construct
the various duals. As these translations are generated by
$$
\pmatrix{e^{\alpha}&0&0\cr
0&e^{-\alpha}&0\cr
0&0&1}\,\qquad  {\rm and} \qquad
\pmatrix{e^{{\beta\over\sqrt{3}}}&0&0\cr
0& e^{{\beta\over\sqrt{3}}}&0\cr
0&0& e^{{-2\beta\over\sqrt{3}}}},
\equation
$$
an easy calcualtion shows that the following $\bR^2$ invariant combinations 
are possible: 
$(J_\mu^3)^2,\,(J_\mu^8)^2,\,J_\mu^1\,J^{2\mu},\,J_\mu^4\,
J^{\mu 5},\,J_\mu^6\,J^{\mu 7}\,.$ Therefore, an arbitrary linear combination 
of these terms could be added to eq.(16) to obtain an 
$SL(3)_L\times\bR_R^2$ symmetric deformed principal sigma model. Note, 
however, that after an appropriate rescaling of $\lambda$ in eq.(16) and the 
coefficients of the linear combination, {\sl any} one of the previous 
invariants can be absorbed into 
the term describing the principal model. Thus the maximum number of free
parameters characterizing the sigma models with $SL(3)_L\times\bR_R^2$ 
symmetry in addition to the coupling, $\lambda$, is four. To reduce this 
number we look for {\sl discrete} subgroups of $SL(3)$ that can be imposed
as symmetries in addition to $\bR^2$. The most natural candidates are the
$\bZ_2$ subgroups generated by the elements:
$$
M_1=\pmatrix{0&1&0\cr
1&0&0\cr
0&0&1},\qquad 
M_2=\pmatrix{1&0&0\cr
0&0&1\cr
0&1&0},\qquad
M_3=\pmatrix{0&0&1\cr
0&1&0\cr
1&0&0},
\equation
$$
and the $\bZ_3$ subgroup generated by
$$
z_1=\pmatrix{0&1&0\cr
0&0&1\cr
1&0&0},\qquad 
z_1^2=z_2=\pmatrix{0&0&1\cr
1&0&0\cr
0&1&0}.
$$
A short calculation, 
using the action of these elements on $G$ ($G\rightarrow a^{-1}Ga$; $a=M_i$, 
$z$, $z^2$) 
and the definition of the 
current, shows that
$$
\alpha (J_\mu^3)^2+\beta (J_\mu^8)^2+\delta J_\mu^1J^{2\mu}
+\gamma (J_\mu^4J^{\mu 5}+J_\mu^6J^{\mu 7})
\equation
$$
is invariant under $M_1$,
$$
\alpha ((J_\mu^3)^2+(J_\mu^8)^2)+\delta J_\mu^6J^{7\mu}
+\gamma (J_\mu^4J^{\mu 5}+J_\mu^1J^{\mu 2})
\equation
$$
is invariant under $M_2$,
$$
\alpha ((J_\mu^3)^2+(J_\mu^8)^2)+\delta J_\mu^4J^{5\mu}
+\gamma (J_\mu^1J^{\mu 2}+J_\mu^6J^{\mu 7})
\equation
$$
is invariant under $M_3$, and 
$$
\alpha ((J_\mu^3)^2+(J_\mu^8)^2)+\beta (J_\mu^4J^{5\mu}
+J_\mu^1J^{\mu 2}+J_\mu^6J^{\mu 7})
\equation
$$
is invariant under $\bZ_3$. Note that requiring the $\bZ_3$ symmetry implies
the various $\bZ_2$ symmetries as well, i.e. choosing eq.(22) guarantees that
the cyclic $\bZ_3$ symmetry is enhanced to ${\rm S}_3$. We mention in passing
that although the action of these discrete subgroups on $J_\mu^a$ is simple,
on the fields, ${\rm F}_i$, they act in general in a rather non-trivial way;
e.g. under $M_1$ the various ${\rm F}_i$'s transform as
$$
\eqalign{
{\rm F}_1&\rightarrow -{\rm F}_1-\ln (1+{\rm F}_3{\rm F}_6),\qquad 
{\rm F}_2\rightarrow {\rm F}_2,\cr
{\rm F}_7&\rightarrow {\rm F}_8,\qquad {\rm F}_8\rightarrow {\rm F}_7,\qquad
{\rm F}_4\rightarrow {{\rm F}_5\over1+{\rm F}_3{\rm F}_6},\cr 
{\rm F}_6&\rightarrow {{\rm F}_3\over1+{\rm F}_3{\rm F}_6},\qquad
{\rm F}_5\rightarrow {\rm F}_4(1+{\rm F}_3{\rm F}_6),\qquad
{\rm F}_3\rightarrow {\rm F}_6(1+{\rm F}_3{\rm F}_6).}
\equation
$$

\underbar{3.2 The simplest deformed principal model and its renormalization}

Using the 
aforementioned rescaling argument to absorb the second term in eq.(22) into
the expression of the principal model we write the Lagrangian of 
the \lq most symmetric' 
deformed $SL(3)$ principal model -- that contain just one parameter in 
addition to $\lambda$ -- as:
$$
{\cal L}=-{1\over2\lambda}\bigl({\rm Tr}J_\mu J^\mu+
2g((J_\mu^3)^2+(J_\mu^8)^2)\bigr)={1\over2\lambda}
g_{ij}({\rm F},g)\d^\mu{\rm F}^i\d_\mu{\rm F}^j.
\equation
$$
Here, by writing the second equality, we emphasize that this model is a purely
metric one, with $g_{ij}$ depending also on the \lq deformation' 
parameter $g$. Since the translations of
${\rm F}_1$ and ${\rm F}_2$ do not commute with the action of ${\rm S}_3$,
the total global symmetry of this Lagrangian is the semidirect product:  
$SL(3)_L\times \bR_R^2\bowtie_s S_3$. 

Explicitely, the non vanishing components of the metric are:
{\small $$
\eqalign{
&g_{11}=-2-2\,g \quad
g_{16}=-2\,\left (1+g\right ){\rm F_3} \quad
g_{17}=\left (1+g\right )\left ({\rm F_4}\,{\rm F_6}\,{\rm F_3}+{\rm F_4}+{\rm F_6}\,{\rm F_5}\right ) \cr
&g_{18}=-\left (1+g\right )\left ({\rm F_3}\,{\rm F_4}+{\rm F_5}\right ) \quad
g_{22}=-2-2\,g \cr
&g_{27}=-\sqrt {3}\left ({\rm F_4}\,{\rm F_6}\,{\rm F_3}-{\rm F_6}\,{\rm F_5}+
{\rm F_4}\right )\left (1+g\right ) \cr
&g_{28}=\sqrt {3}\left ({\rm F_3}\,{\rm F_4}-{\rm F_5}\right )\left (1+g\right ) \quad
g_{36} =-1 \quad
g_{ 47} =-1-{\rm F_6}\,{\rm F_3} \quad
g_{ 48}= {\rm F_3} \quad
g_{57}= {\rm F_6} \quad \cr
&g_{6 6} =-2\,g{\rm F_3}^{2} \quad
g_{67}=g{\rm F_3}\,\left ({\rm F_4}\,{\rm F_6}\,{\rm F_3}+{\rm F_4}+{\rm F_6}\,{\rm F_5}\right ) \quad
g_{6 8}=-g{\rm F_3}\,\left ({\rm F_3}\,{\rm F_4}+{\rm F_5}\right ) \cr
&g_{7 7 }=-2\,g\left (-{\rm F_5}\,{\rm F_3}\,{\rm F_4}\,{\rm F_6}^{2}+{\rm F_4}^{2}+2
\,{\rm F_3}\,{\rm F_4}^{2}{\rm F_6}+{\rm F_3}^{2}{\rm F_4}^{2}{\rm F_6}^{2}+
{\rm F_5}^{2}{\rm F_6}^{2}-{\rm F_5}\,{\rm F_4}\,{\rm F_6}\right ) \cr
&g_{7 8}=g\left (2\,{\rm F_3}^{2}{\rm F_4}^{2}{\rm F_6}+2\,{\rm F_5}^{2}{\rm F_6}-2
\,{\rm F_5}\,{\rm F_3}\,{\rm F_4}\,{\rm F_6}-{\rm F_5}\,{\rm F_4}+2\,{\rm F_3}\,{\rm F_4}^{2}\right ) \cr
&g_{8 5}=-1 \quad g_{8\,8}=-2\,g\left ({\rm F_3}^{2}{\rm F_4}^{2}+{\rm F_5}^{2}-{\rm F_3}\,{\rm F_5}\,{\rm F_4}\right )\,.
}.
\equation
$$} 
Now we discuss briefly the implementation of the coupling constant
renormalization procedure for this model.

Computing the counterterms -- using eq.(5-7) -- for the Lagrangian in 
eq.(24) reveals that both $\Sigma_1$ and $\Sigma_2$ preserve the structure
of the metric, eq.(25): all the non vanishing terms of $\Sigma_1$ and 
$\Sigma_2$ have the same polynomial forms in ${\rm F}_i$ and 
$\d_\mu{\rm F}_i$ as in eq.(24), the only change is in the coefficients 
of the various terms. Therefore we try to abstract the renormalization of
$\lambda$ and $g$ by assuming an ordinary wave function renormalization
for the ${\rm F}_i$ fields, i.e. we substitute 
$$
\eqalign{
F_i^0=Z_i(g,\lambda)\,F_i\quad i=\overline{1,8}\,,
\quad g_0=Z_g(g,\lambda)\,g\,, \quad
\lambda^0=Z_\lambda(g,\lambda)\,\lambda\,,
}
\equation
$$
with
$$
\eqalign{
&Z_i(g,\lambda)=1+{1\over\epsilon}
\Bigl({y^{(1)}_i(g) \lambda\over\pi}+{y^{(2)}_i(g)\lambda^2
\over8\pi^2}+\ldots
\Bigr)+\ldots \,,\cr 
&Z_g(g,\lambda)=1+{y_g(g ,\lambda)\over\epsilon}+\ldots \,,\quad 
Z_\lambda(g,\lambda)=1+{y_\lambda(g ,\lambda)\over\epsilon}+\ldots \,,
}
\equation
$$
into eq.(9, 10) in both the one and two loop order.
(In eq.(27) $y_\lambda$ and $y_g$ are given by eq.(8), 
after the obvious $\zeta_i(x)
\rightarrow y_\lambda^i(g)$ and $x_i(x)\rightarrow y_g^i(g)$ $i=1,2$ 
changes).  Eq.(9) yields, in 
both cases, $66$ linear equations for the $10$ unkowns $y^{(j)}_i(g),\ 
i=\overline{1,8}$, $y_\lambda ^j(g)$, $y_g^j(g)$; however as a result of
the high degree of symmetry of eq.(24) and the manifest target space 
covariance of the background field method this system is solvable. In fact
$y_\lambda ^j (g)$ and $y_g^j(g)$ are determined uniquely, while there is 
a two parameter freedom in the choice of the 
$8$ wave function renormalization constants $y^{(j)}_i(g)$. This freedom 
reflects the following two independent scaling invariances of the Lagrangian,
eq.(24):
$$
\eqalign{
&F_1\longrightarrow\pmatrix{1 \cr 1}\,F_1\quad
F_2\longrightarrow\pmatrix{1 \cr 1}\,F_2\quad
F_3\longrightarrow\pmatrix{e^\gamma \cr e^\delta}\,F_3\quad
F_4\longrightarrow\pmatrix{e^{-\gamma} \cr 1}\,F_4 \cr
&F_5\longrightarrow\pmatrix{1 \cr e^\delta}\,F_5\quad 
F_6\longrightarrow\pmatrix{e^{-\gamma} \cr e^{-\delta}}\,F_6\quad
F_7\longrightarrow\pmatrix{e^\gamma \cr 1}\,F_7\quad
F_8\longrightarrow\pmatrix{1 \cr e^{-\delta}}\,F_8.
}
\equation
$$
In one loop 
the explicit form of the coupling constant and deformation parameter's 
renormalization constants are:
$$
y_g(g,\lambda)
={5\,\lambda\,\left (1+g\right )\over4\,\pi}\,,\qquad
y_\lambda (g,\lambda)
={\left (2\,g-3\right )\,\lambda\over4\,\pi}\,,
\equation
$$
while in two loops:
$$
\eqalign{
&y_g(g,\lambda)=-{\lambda (1+g)(10g\lambda-13\lambda
-40\pi )\over32\pi^2},\cr
&y_\lambda(g,\lambda)
=-{\lambda\,\left (48\,\pi+26\,\lambda\, g^{2}-32\,\pi\,g-11\,g\,\lambda+
9\,\lambda\right )\over64
\,\pi^{2}},}
\equation
$$
was obtained. Using these expressions in eq.(11) 
leads to the following one and two loop 
$\beta$ functions:
$$
\eqalign{
&\beta_\lambda^{(1)}={\lambda^{2}\,\left (2\,g-3\right )\over4\,\pi}\,, 
\qquad\cr
&\beta_g^{(1)}={5\,\lambda\,\left (1+g\right )\,g\over4\,\pi}\,,
}  
\equation
$$
$$
\eqalign{
&\beta_\lambda^{(2)}={\lambda^{2}\left (2\,g-3\right )\over4\,\pi}+
{\lambda^{3} 
\left(26\,g^{2}-11\,g+9\right )\over32 \,\pi^2}\,, \cr
&\beta_g^{(2)}={5\,\lambda\left (1+g\right )g\over4\,\pi}+
{\lambda^2 g(10\,g^2-3\,g-13)\over16\pi^2}\,.}
\equation
$$
It is easy to see that the $g=0$ resp. the $g=-1$ lines are fixed lines 
under the renormalization group in the $(\lambda ,g)$ plane. On the $g=0$ line
(that corresponds to the principal model) $\beta_\lambda^{(2)}(\lambda)$ 
indeed
reduces to the well known 
$\beta$ function of the $SU(3)$ principal model [14] if we take 
into account the normalization of $\lambda$ implied by eq.(16). 
Looking at eq.s (24,25) we conclude that the $g=-1$
fixed line -- that also describes an asymptotically free model -- can be 
interpreted as a sort of "coset model" where the $J_\mu^3$, $J_\mu^8$ 
current components -- or the 
${\rm F}_1$ and ${\rm F}_2$ fields, decouple completely. Away from these lines 
in the ($\lambda\ge0$, $g<0$) quarter of the ($\lambda$,$g$) plane
the renorm trajectories
run into $\lambda=0$, $g=-1$; while for $g>0$ they run to infinity. This
implies that the $g=0$ fixed line corresponding to the principal
$\sigma$-model is `unstable' under the deformation.     

As the metric, eq.(25), 
 is independent of ${\rm F}_1$, ${\rm F}_2$, ${\rm F}_7$ and
${\rm F}_8$, the simplest deformed $SL(3)$ principal 
model has four Abelian isometries 
(corresponding to the translations of these fields), that can be used to 
construct various dual models. Of these four possibilities we are 
concerned here only with the ones using the ${\rm F}_1$ and ${\rm F}_2$
translations as  
$g_{11}$ and $g_{22}$ are {\sl constants}, while $g_{77}$ and $g_{88}$ depend
on the other ${\rm F}$'s. In the sequel we call \lq ${\rm F}_1$ (resp. 
${\rm F}_2$) dual' of any \lq original' model the dual models obtained
by using the ${\rm F}_1$ (resp. ${\rm F}_2$) translations to generate them.  

\chapter{The ${\rm F}_1$ dual of the simplest deformed principal model}

The Lagrangian of 
the ${\rm F}_1$ dual of the simplest deformed principal model, ${\cal L}^d$,  
is easily 
obtained from using eq.(3) and eq.(24-25). Interestingly this dual model
contains a non trivial torsion term, in addition to a somewhat different
metric. Appropriately rescaling the field dual to ${\rm F}_1$ and  
-- with a slight abuse of notation -- still denoting it by ${\rm F}_1$,    
the non vanishing components of $b_{ij}$ and $g_{ij}$ are:
{\small
$$
\eqalign{ 
&b_{16}=\sqrt{1+g}{\rm F_3}, \quad
b_{17}=-{1\over2}\sqrt{1+g}\left ({\rm F_4}\,{\rm F_6}\,{\rm F_3}+{\rm F_4}
+{\rm F_6}\,{\rm F_5}\right ), \cr
&b_{18}={1\over2}\sqrt{1+g}\left ({\rm F_3}\,{\rm F_4}+{\rm F_5}\right ),}
\equation
$$
$$
\eqalign{
&g_{11}=-{1\over2} \quad
g_{22}=-2-2\,g \quad
g_{27}=-\sqrt {3}\left ({\rm F_4}\,{\rm F_6}\,{\rm F_3}-{\rm F_6}\,{\rm F_5}+
{\rm F_4}\right )\left (1+g\right ) \cr
&g_{28}=\sqrt {3}\left ({\rm F_3}\,{\rm F_4}-{\rm F_5}\right )
\left (1+g\right ) \quad
g_{36} =-1 \quad
g_{ 47} =-1-{\rm F_6}\,{\rm F_3} \quad
g_{ 48}= {\rm F_3} \quad
g_{57}= {\rm F_6} \quad \cr
&g_{6 6} =2{\rm F_3}^{2} \quad
g_{67}=-{\rm F_3}\,\left ({\rm F_4}\,{\rm F_6}\,{\rm F_3}+{\rm F_4}+{\rm F_6}\,{\rm F_5}\right ) \quad
g_{68}={\rm F_3}\,\left ({\rm F_3}\,{\rm F_4}+{\rm F_5}\right ) \cr
&g_{7 7 }=(3g+1){\rm F_5}\,{\rm F_4}\,{\rm F_6}({\rm F_6}{\rm F_3}+1)
+\cr &{1-3g\over2}({\rm F_4}^{2}+2
\,{\rm F_3}\,{\rm F_4}^{2}{\rm F_6}+{\rm F_3}^{2}{\rm F_4}^{2}{\rm F_6}^{2}+
{\rm F_5}^{2}{\rm F_6}^{2})\cr
&g_{7 8}={1-3g\over2}(2\,{\rm F_3}{\rm F_4}^{2}{\rm F_6}+
{\rm F_3}^2{\rm F_4}^{2}{\rm F_6}+
\,{\rm F_5}^{2}{\rm F_6}^2+{\rm F_4}^2)\cr &+(3g+1)
\,{\rm F_5}{\rm F_4}{\rm F_6}({\rm F_3}{\rm F_6}+1)\cr
&g_{8 5}=-1 \quad 
g_{8\,8}={1\over2}(1-3g)({\rm F_3}^{2}{\rm F_4}^{2}+{\rm F_5}^{2})
+(3g+1){\rm F_3}\,{\rm F_5}\,{\rm F_4}.
}
\equation
$$}
Note that the global symmetries of this dual model consist of the full 
$SL(3)_L$ transformations (that include among others the translations of 
${\rm F_7}$ and ${\rm F_8}$ as well as the scaling transformations
of eq.(28)) and the translations of ${\rm F_1}$ and ${\rm F_2}$. The
discrete ${\rm S}_3$ part of the original symmetry group is missing: the 
${\rm F_1}$ translation used to construct the dual does not commute with it. 

Applying the coupling constant renormalization procedure in the same way as
for the purely metric \lq original' model leads, {\sl in the one loop order},
to the same $\beta$-functions, eq.(31). However for two loops the   
situation changes 
dramatically: the structure of the Lagrangian is not reproduced by the 
countertems as completely 
new terms appear in the symmetric part of $\Sigma_2$:  
$$
\alpha_0({\rm F_4}\,{\rm F_6}\,\d_\mu{\rm F_3}\d^\mu{\rm F_7}-
{\rm F_4}\,\d_\mu{\rm F_3}\d^\mu{\rm F_8}+
{\rm F_5}\,\d_\mu{\rm F_6}\d^\mu{\rm F_7}),
\equation
$$
where $\alpha_0=3\lambda^2(1+g)(5-3g)/(32\pi^2)$. These terms are new, since
in the metric, eq.(34), $g_{37}$ and $g_{38}$ vanish and $g_{67}$ contains 
no single ${\rm F_5}$. We note, that   
these new terms are manifestly invariant under
the four translational and two scaling (eq.(28)) invariances of 
${\cal L}^d$, a fact reflecting that 
dimensional regularization preserves the
global symmetries.

To establish the two loop equivalence of this dual model to eq.(24) we 
must be able, as a first step, to extract from $\Sigma_2$ the two loop 
renormalization constants of $\lambda$ and $g$. If this is completed, then
we have to inquire whether the $\beta$-functions obtained from these 
renormalization constants are really equivalent -- possibly after an
appropriate change of the scheme [7] -- to those in eq.(32). 

To extract the renormalization constants of $\lambda$ and $g$ we look for
reparametrizations of ${\rm F_i}$'s that would account for the new terms
in eq.(35) while preserving the manifest 
symmetries (the four translations and 
the two scaling invariances in eq.(28)). It is easy to see, that 
$$   
{\rm F_5}\rightarrow{\rm F_5}+\alpha_0{\rm F_3}{\rm F_4}/\epsilon
\equation
$$  
accounts for the the first two terms in eq.(35) while it preserves the
polynomial form of the other metric and torsion components in eq.(33,34).
However we proved that there does not exist such reparametrization of the
${\rm F_i}$'s that

\item{$\bullet$} would preserve the manifest symmetries,

\item{$\bullet$} would produce the 
${\rm F_5}\,\d_\mu{\rm F_6}\d^\mu{\rm F_7}$ term,

\item{$\bullet$} would not change the polynomial structure of the other 
metric and torsion components.

\noindent The proof is completed by writing down the most 
general reparametrization,
compatible with the first two requirements and by showing that the last
requirement eliminates all potential free terms in it.

Therefore, at two loops, within the subspace spanned by $\lambda$ and $g$, 
the ${\rm F}_1$ dual of eq. (24) is not renormalizable in the ordinary, 
field theoretical sense, and the $2$ loop $\beta$-functions cannot be 
extracted. Thus, naively, one is tempted to conclude that the ${\rm F_1}$ 
dual cannot be equivalent to the original model. However, since this non
renormalizability follows from the appearance of new terms, not accounted for
by reparametrizations, one has to understand this phenomenon better, before
saying anything definit about the quantum equivalence between eq.(24) and 
${\cal L}^d$. There are two points one has to take into account: the first -- 
as we emphasized earlier -- is that ${\cal L}^d$ has fewer symmetries than 
${\cal L}$, eq.(24), thus the appearance of new terms may be attributed to
this symmetry reduction. The second point worth emphasizing is that 
describing ${\cal L}$ in terms of $\lambda$ and $g$ is not renorm invariant; 
for this one should use a quantity, $C=C(\lambda ,g)$, depending on $\lambda$ 
and $g$, which is constant along the renorm trajectories: ${\d C\over
\d\lambda}\beta_\lambda(\lambda ,g)+{\d C\over\d g}\beta_g(\lambda ,g)
=0$, and the \lq physical' $\beta$-function is obtained from $\beta_\lambda
(\lambda ,g)$ by expressing in it $g$ in terms of $C$. Therefore, putting 
these two observations together, it is conceivable, that for ${\cal L}^d$, 
working in a larger parameter space (i.e. having more parameters than just
$\lambda ,\ g$), and requireing the renorm invariants to be constants even at 
two loops forces us out of the $(\lambda ,g)$ subspace, yet when the 
parameters are expressed in terms of the invariants, the physical $\beta^d$ 
still coincides with that of the original model.
  
To close this section we mention that we found qualitatively  the same 
behaviour for the ${\rm F}_2$ dual of the simplest deformed principal 
sigma model. The global symmetries of this model contain not only $SL(3)_L
\times{\bf{\rm R^2}}$ but also 
 the ${\bf{\rm Z_2}}$ 
subgroup of ${\rm S_3}$ generated by $M_1$, eq.(18), as the action of $M_1$ 
on the fields, eq.(23), commutes with the translations of ${\rm F}_2$. 
Nevertheless, while in one loop we found the $\beta$-functions, eq.(31), at 
two loops new terms, similar to those in eq.(35), arose in the counterterms,  
and they destroyed the renormalizabilty of this model in the same way as in
the case of the ${\rm F}_1$ dual.

\chapter{Models with more parameters}

\underbar{5.1 Derivation of the $\beta$ functions}

A systematic way to introduce more parameters into the Lagrangian, 
${\cal L}^d$, of the previous section, is to construct it as the 
dual of the less symmetric deformed
$SL(3)$ principal models described in sect.3. Of the many possibilities 
discussed there 
we choose the \lq $M_1$ invariant', whose Lagrangian can be written as (see
eq.19):
$$\eqalign{
{\cal L}&=-{1\over2\lambda}\bigl({\rm Tr}J_\mu J^\mu-
4b(J_\mu^3)^2-4c(J_\mu^8)^2-dJ_\mu^1J^{\mu 2}\bigr)\cr 
&={1\over2\lambda}
g_{ij}({\rm F},b,c,d)\d^\mu{\rm F}^i\d_\mu{\rm F}^j,} 
\equation
$$  
since this is the only one where the counterterms reproduce the structure
of the Lagrangian thus 
the two loop renormalization requires  
no re\-pa\-ra\-met\-ri\-za\-tion of 
the ${\rm F_i}$'s. The explicit form of the metric
$$
\eqalign{
&g_{11}=-2(1-2b) \quad
g_{16}=-2(1-2b){\rm F_3}\quad
g_{17}=(1-2b)\left ({\rm F_4}+{\rm F_3}\,{\rm F_6}
\,{\rm F_4}+{\rm F_6}\,{\rm F_5}\right )\cr
&g_{18}=-(1-2b)\left ({\rm F_3}\,{\rm F_4}+{\rm F_5
}\right )\quad 
g_{27}=-\sqrt {3}\left ({\rm F_4}-{\rm F_6}\,{\rm F_5}+
{\rm F_3}\,{\rm F_6
}\,{\rm F_4}\right ) (1-2c)\cr
&g_{22}=4c-2\quad g_{28}=\sqrt {3}\left (-{\rm F_5}+{\rm F_3}\,{\rm F_4}\right )
(1-2c)\cr
&g_{36} =-1+{d\over2}\quad
g_{37} =-{d\over2}{\rm F_6}\,{\rm F_4}\quad
g_{38} ={d\over2}{\rm F_4}\quad
g_{ 47} =-1-{\rm F_3}\,{\rm F_6}\quad
g_{58}=-1\cr
&g_{66} =(4b-d){\rm F_3}^{2}\quad
g_{67}=({d\over2}-2b)\left ({\rm F_3}^{2}{\rm F_6}\,{\rm F_4}+
{\rm F_3}\,{\rm F_6}\,{\rm F_5}\right )+{d\over2}{\rm
F_5}+(1-2b){\rm F_3}\,{\rm F_4} \cr
&g_{7 7 }=-x\,\left ({\rm F_5}^{2}{\rm F_6}^{2}+{\rm F_3}^{2}{\rm F_6}^{2}{\rm F_4}^{2}+2{\rm F_6}\,{\rm F_4}^{2}
{\rm F_3}+{\rm F_4}^{2} \right )+y\left ({\rm F_6}\,{\rm F_4}\,{\rm F_5}+{\rm F_3}\,{\rm F_6}^{2}{\rm F_4}\,{\rm 
F_5}\right )\cr
&g_{ 48}={\rm F_3} \quad
g_{7 8}=x\,\left ({\rm F_3}^{2}{\rm F_4}^{2}{\rm F_6}+{\rm F_3}\,{\rm F_4}^{2}+{\rm F_6}\,{\rm F_5}^{2}\right
)-y\,\left ({1\over2}\,{\rm F_4}\,{\rm F_5}+{\rm F_3}\,{\rm F_6}\,{\rm F_4}\,{\rm F_5}\right )\cr
&g_{57}={\rm F_6}\quad
g_{68}=(2b-{d\over2})\left ( {\rm F_4}\,{\rm F_3}^{2}+{\rm F_5}\,{\rm F_3} \right )\quad
g_{8 8}=x\,\left ({\rm F_5}^{2}+{\rm F_4}^{2}{\rm F_3}^{2}\right )+
y\,{\rm F_5}\,{\rm F_3}\,{\rm F_4}\,,
}
\equation
$$
where $x\equiv -b-3c$ and $y\equiv 2b-d-6c$, reveals that the terms appearing
in eq.(35) are already present in the original model. The formulae of the
duality transformations, eq.(3), guarantee that these terms are also there 
in the Lagrangian of both the ${\rm F_1}$ and the ${\rm F_2}$ duals of 
eq.(37). Setting $d\equiv 0$ and $b=c=-g/2$ in eq.(37) yields the Lagrangian 
of the simplest deformed principal  model, eq.(24).

We applied the coupling constant renormalization procedure for the $4$ 
parameter purely metric sigma model, eq.(37), as well as for its ${\rm F_1}$ 
and ${\rm F_2}$ duals. In all cases it turned out that in this larger 
parameter space the $\Sigma_1$ and 
$\Sigma_2$ counterterms reproduce the structure of the corresponding 
Lagrangians, and up to two loops each of the three models proved to be 
renormalizable. The one loop $\beta$-functions
$$
\eqalign{
\beta_\lambda^1&=-{\lambda^{2}\left (d+6\,c+6+
2\,b\right )\over8\,\pi},\cr
\beta_b^1&=
{\lambda\left (d^{2}+d+3\,dc-bd-12\,bc-12\,b^{2}+8-4\,b\right )\over16\,
\pi }+{8\lambda b(1-b)-2\lambda\over2\,\pi 
\left (d-2\right )^{2}}\cr
&+{\lambda (1-2\,b)\over2\,\pi\left (d-2\right )}
,\cr
\beta_c^1&=
-{\lambda\left (2\,cd-d+36\,c^{2}-2\,b-18\,c+4\,bc\right )\over16\,\pi},\cr
\beta_d^1&=\lambda\left (-{d^{2}+3\,cd+bd+4-6\,c-2\,b\over4\,\pi}+
{4\,b-2\over\pi\,\left (d-2\right )}\right ),}
\equation
$$
coincide in all three cases, while 
the case by case different, and somewhat more
complicated two  loop expressions are given in Appendix A. (In a form 
 appropriately transformed to describe the evolution of the parameters 
introduced in eq.(42)). 
  
These $\beta$-functions contain 
some terms depending in a non polynomial way on the parameter $d$. This 
reflects our assumption that the $y_K(\lambda ,b,c,d)$ 
residues of the simple poles in the coupling and  wave function 
renormalization constants, $Z_K=1+y_k(\lambda ,b,c,d)/\epsilon+...$, 
($K=\overline{1,8}$, $\lambda ,\ b,\ c,\ d$) depend polynomially only on the
coupling constant, $\lambda$, while they may depend in an arbitrary way 
on $b,\ c,\ d$, as in these parameters we do not expand anything. Note also
that for $d\equiv 0$ and $b=c=-g/2$ the expressions  in eq.(39) become 
equivalent to the \lq universal' 2 parameter one loop $\beta$-functions, 
eq.(31), while the metric 2 
loop $\beta$-functions in Appendix A  reduce  to eq.(32).

Now that we succesfully obtained the $\beta$-functions up to two loops 
not only for  the original  model, eq.(37), but also for its two duals, and 
these $\beta$-functions are different from each other, we can discuss the 
quantum equivalence among them. To this end -- as discussed at length in 
ref.[7] -- first one should determine the renorm invariants 
$N(\lambda ,b,c,d)$, 
$$ 
{\d N\over
\d\lambda}\beta_\lambda+{\d N\over\d b}\beta_b+{\d N\over
\d c}\beta_c+{\d N\over\d d}\beta_d=0,
\equation
$$
characterizing the trajectories for each model in question. Eq.(40) is a 
homogeneous, linear partial differential equation, thus by general theorems 
it admits -- at least locally -- , in the present case, $3$ independent 
first integrals $N_i(\lambda ,b,c,d)$ $i=1,2,3$. Then, for each model, one 
should solve (of course only perturbatively in $\lambda$) the 
$N_i(\lambda ,b,c,d)=N_i^0$ equations for $b,\ c,\ d$ and use the $b=
b(\lambda ,N_i^0)$ etc. expressions in the corresponding $\beta_\lambda$ to 
obtain the \lq physical' $\beta$-functions. Then the models may be equivalent 
if the various physical $\beta$-functions coincide for the same values of 
$N_i^0$.       

The actual integration of eq.(40) is a formidable task that greatly exceeds
our capabilities. This is so even if we realise that the first step would
be to determine $N_i(\lambda ,b,c,d)$ for
the one loop $\beta$-functions, eq.(39), common for all three models, and the
two loop corrections could be obtained perturbatively. What we can do is
a sort of \lq fixed point analysis': 
we linearize (in $b,\ c,\ d$) the various $\beta$-functions around 
the particular $b,\ c,\ d$
values (\lq fixed points'), satisfying
$$
\beta_b=\beta_c=\beta_d=0  ,
\equation
$$
and use them in eq.(40). 

Though the one loop $\beta$-functions, eq.(39) allow a number of 
real fixed points
$$\eqalign{
&(b=c=d=0),\quad (b=c=1/2\ d=0),\quad (b=c=1/2\ d=-2),\cr
&(b=-1/2\ c=1/2\ d=-2),\quad (b=3/10\ c=1/2\ d=6/5),}
$$
it is only the $(b=c=1/2\ d=0)$ one that makes all three two loop sets of
$\beta_b$, $\beta_c$, and $\beta_d$ vanish. Therefore we carry out the 
 the aforementioned analysis for this  \lq fixed point', which, incidentally,
describes the coset model where the $J_\mu^3$ and $J_\mu^8$ current 
components decouple.  The analysis becomes tractable by introducing the
$B,\ C,\ D$ parameters:
$$
b=-{1\over2}(B(D+1)+D),\qquad c=-{1\over2}C-{1\over6}D,\qquad 
d=-2D,
\equation
$$
in terms of which the one loop $\beta$-functions simplify:
$$
\eqalign{
&\beta^{(1)}_\lambda= {\lambda^{2}\left (B+BD+4\,D-
6+3\,C\right )\over8\,\pi},\quad 
\beta_B^{(1)}= {B\,\lambda\,\left (5+2\,D+D^{2}\right )
\left (B+1\right )\over4\,\pi\,
\left (1+D\right )},\cr
&\beta_C^{(1)}= {\lambda\left (6\,B+2\,BD+3\,BDC+3\,BC+24\,C+
27\,C^{2}+24\,DC+16\,D
\right )\over24\,\pi},\cr
&\beta_D^{(1)}= {\lambda\left (BD^{2}+6\,D^{2}+2\,D-3\,B+2\,BD+
3\,C+3\,DC\right )\over8
\,\pi},}
\equation
$$
and also the two loop expressions acquire a somewhat less complicated 
form. In terms of these new parameters $(B=-1,\ C=-1,\ D=0)$ is 
the fixed  point describing  
the coset model, while the subspace corresponding to the most symmetric 
deformed principal model (and its duals) is obtained by the substitution 
$B\equiv C\equiv g$, $D\equiv 0$.  

\underbar{5.2 Analysis around the \lq fixed point'}

We linearize the three one and two loop $\beta$ functions $\beta_B$, 
$\beta_C$, $\beta_D$, by writing 
$$
B=-1+\hat{b}, \qquad C=-1+\hat{c},\qquad D=0+\hat{d},
\equation
$$
for both the metric model, eq.(37), as well as 
 for the ${\rm F}_1$ and ${\rm F}_2$ duals. (We 
emphasize, that the linearization does not effect the $\lambda$ dependence).
Assembling $[\hat{b},\hat{c},\hat{d}]$ into the components of a vector, the
right hand sides of eq.(43) and  
(A.2-A.4), (A.6-A.8), (A.10-A.12),
yield in each case a $3\times 3$ matrix acting on 
$[\hat{b},\hat{c},\hat{d}]$, whose eigenvalues and 
eigenvectors play an important role in the following. For the metric model 
the eigenvalues and the corresponding eigenvectors of this matrix are:
$$\eqalign{
&-{\lambda\over16\pi^2}(20\pi +23\lambda )\qquad [1,0,{3\over7}+
{345\over196\pi}\lambda ],\cr
&-{\lambda\over16\pi^2}(20\pi +23\lambda )\qquad [0,1,-{3\over7}-
{345\over196\pi}\lambda ],\cr
&-{\lambda\over16\pi^2}(8\pi +31\lambda )\qquad \  [0,1,-3],}
\equation
$$
while for the the ${\rm F}_1$ and ${\rm F}_2$ duals the results coincide:
$$\eqalign{
&-{\lambda\over16\pi^2}(20\pi +23\lambda )\qquad [1-{2\lambda\over\pi},1,
0],\cr
&-{\lambda\over16\pi^2}(20\pi +23\lambda )\qquad [{7\over3}-{41\over4\pi}
\lambda ,0,1],\cr
&-{\lambda\over16\pi^2}(8\pi +31\lambda )\qquad \  [0,1,-3].}
\equation
$$  
In these expressions the terms linear (constant) in $\lambda$ represent the
one loop results, while the quadratic (linear) ones describe the two loop
corrections. Note that up to this order the three eigenvalues for the metric
and the ${\rm F}_1$, ${\rm F}_2$ duals are the same in spite of the different
$\beta$ functions. However the two loop eigenvectors in case of the 
${\rm F}_1$/${\rm F}_2$ duals are different from the metric ones.

Keeping only the constant and linear terms of $\hat{b}$, $\hat{c}$,
and $\hat{d}$ also in $\beta_\lambda$ simplifies the one loop renormalization
equations, eq.(43): 
$$
{d\lambda\over dt}=-{\lambda^2\over8\pi}\bigl(\sum\limits_ja_jv_j+L\bigr);
\qquad {dv_i\over dt}={\lambda\over8\pi}\Delta_iv_i,
\equation
$$
where ${\lambda\over8\pi}\Delta_i$ ($i=1,2,3$) denote the three ($1$ loop)
eigenvalues appearing in eq.(45,46), $v_i$ stand for the components of the 
corresponding eigenvectors, and in the last equation there is no summation 
over $i$. These equations admit the following renorm invariants:
$$
N_i={\lambda\over v_i^{-L/\Delta_i}}\exp\bigl(\sum{a_k\over\Delta_k}v_k\bigr),
\qquad i=1,2,3.
\equation
$$ 
Nevertheless this is not yet the end of the story. The point is 
that expressing $v_i=v_i(\lambda , N_j)$ from eq.(48) makes sense in 
perturbation theory only if $v_i$ stay small for $\lambda\rightarrow 0$; 
i.e. if 
$$
0<-{L\over\Delta_i}\le 1.
\equation
$$
Therefore our analysis yields acceptable results only if these conditions
are satisfied; it is easy to see that eq.(49) holds for the two 
\underbar{coinciding} eigenvalues in eq.(45,46), (as $L=10$ and $\Delta_{1,2}
=-10$),  
while it is not satisfied for the third eigenvalue. Therefore our perturbative
results are valid only in the $[\hat{b},\hat{c},{3\over7}(\hat{b}-\hat{c})]$ 
($[\hat{c}+{7\over3}\hat{d},\hat{c},\hat{d}]$) subspaces of the three 
dimensional space spanned by $\hat{b}$, $\hat{c}$, $\hat{d}$ for the metric 
and the ${\rm F}_1$/${\rm F}_2$ dual models respectively.

The fact that according to eq.(45,46) these \lq\lq perturbative" subspaces 
change little (to $[\hat{c}+{7\over3}\hat{d}-{\lambda\over\pi}(2\hat{c}+
{41\over4}\hat{d}),\hat{c},\hat{d}]$ for the ${\rm F}_1$/${\rm F}_2$ dual) 
or none ($[\hat{b},\hat{c},{3\over7}(\hat{b}-\hat{c})]$ stays for the metric 
model) in the two loop order already indicates, that they are indeed 
appropriate to describe the perturbative evolution of these models. To prove 
this more rigorously in Appendix B we show that in these subspaces the two 
loop corrections to $N_i$ are indeed of higher order in $\lambda$. 

In the perturbative subspaces -- exploiting the degeneracy of the 
corresponding eigenvalues -- the $\lambda$ dependence of the remaining two
parameters becomes $p_i\sim {\lambda\over N_i}+o(\lambda^2)$, where 
$p_1=\hat{c}$, $p_2=\hat{d}$ as they effectively arise from the one 
loop results. Since the
two loop ($o(\lambda^3)$) corrections to $\beta_\lambda$ come in the 
form $\lambda^3\otimes (B=C=-1,\ D=0)$fixed point $+$ $\lambda^2\otimes o(
\lambda)$corrections of the one loop terms; and the one loop terms are common
for all three models, the \lq\lq physical" $\beta_\lambda$-s of the various
models coincide if the $\lambda^3\otimes $fixed point pieces coincide. 
Looking at the explicit expressions in Appendix A reveals that this is indeed 
the case; i.e. in the perturbative subspaces, up to two loops, the three models
may indeed be equivalent.

The two loop form of the perturbative subspace of the 
${\rm F}_1$/${\rm F}_2$ duals, 
$[\hat{c}+{7\over3}\hat{d}-{\lambda\over\pi}(2\hat{c}+
{41\over4}\hat{d}),\hat{c},\hat{d}]$, as compared to the metric one  
$[\hat{b},\hat{c},{3\over7}(\hat{b}-\hat{c})]$, also explaines why we needed
to enlarge the parameter space of the simplest deformed principal model, 
which corresponds to $\hat{b}\equiv\hat{c}$, $\hat{d}\equiv0$: while this
is a consistent subspace of the metric two loop perturbative subspace it 
is not consistent with that of the ${\rm F}_1$/${\rm F}_2$ duals. Indeed, for 
the latter case, even if we put $\hat{d}=0$ in the one loop order, to have
$\hat{b}=\hat{c}$ in two loops we need a non vanishing 
$\hat{d}$ of $o(\lambda)$. The 
interpretation of this is clear: since the dual models have less symmetry 
than the simplest deformed principal model there is nothing that would 
prohibit the perturbative generation of the coupling described by $\hat{d}$. 
Nevertheless -- at least when we treat the deviation of $g\equiv B\equiv C$ 
from $-1$, $g=\hat{c}-1$, perturbatively, -- our previous findings about the 
coinciding physical $\beta$ functions for the metric and the ${\rm F}_1$/${
\rm F}_2$ dual models indicate, that the simplest deformed principal model
and its ${\rm F}_1$/${\rm F}_2$ duals in the extended parameter space may, 
indeed, be physically equivalent. 

\chapter{Summary and conclusions}
\bigskip

In this paper we investigated the quantum equivalence between various 
deformed $SL(3)$ principal models and their different duals in the two loop
order of perturbation theory. The investigation is based on extracting and 
comparing various $\beta$ functions of the original and dual models.

To keep the number of parameters that require renormalization 
small, as a first step, we 
determined the Lagrangians of those deformed $SL(3)$ principal models that
admit $SL(3)_L\times R^2\times$some discrete subgroup of $SL(3)$ as global
symmetries. Then we carried out the coupling constant renormalization program
and determined the various $\beta$ functions for the simplest (i.e. most 
symmetric) deformed $SL(3)$ principal model as well as for the more general 
case with $SL(3)_L\times R^2\times Z_2$ symmetry.

We considered only those duals of these sigma models (we called them ${\rm 
F}_1$ and ${\rm F}_2$ duals) that are obtained by a very simple Abelian 
duality transformation when the $g_{00}$ component of the metric is constant. 
This choice is motivated -- besides beeing obviously the simplest possibity --
also by the following: the question of quantum equivalence between the 
original and dual models is basically a two loop problem, as up to one loop 
the equivalence is shown in general in ref.[8]. However to determine the
two loop counterterms of the duals obtained by using a non constant $g_{00}$ 
exceeded greatly the power of our computers. The case of constant $g_{00}$ 
is also favoured by the fact, that in this case the standard (functional 
integral based) derivation of the duality transformations amounts to a 
standard Gaussian integration, thus no problems are expected with the quantum
equivalence of the two models. 

Of the counterterms of the ${\rm 
F}_1$ and ${\rm F}_2$ duals of the simplest 
deformed $SL(3)$ principal model we found that in the two loop order new 
terms appeared that could not be accounted for by some field redefinition. 
Therefore, naively, these duals seem to be non renormalizable in the ordinary,
field theoretic sense. However, since for these dual models some of the 
explicit discrete symmetries of the simplest 
deformed $SL(3)$ principal model are absent, we conjectured, that the 
appearance of the new terms may be the consequence of this symmetry reduction. 
Thus working in a larger parameter space the renormalizability of the dual 
models would be restored and expressing everything in terms of renorm 
invariants, the original and dual $\beta$ functions could still coincide.

We verified explicitly, without any approximation, the restoration of the 
renormalization of the dual models in an appropriately chosen larger parameter 
space. However we could verify the equality of the various \lq physical' 
$\beta$ functions only by treating the deviations of these parameters from 
certain \lq fixed point values' perturbatively. Nevertheless these are 
certainly consistent with the expectation that -- in case of a constant 
$g_{00}$ -- the equivalence of the dually related sigma models also involves 
a change of the renormalization scheme.

\medskip 

We close the paper by an additional  
-- not entirely unrelated -- speculation about the 
origin of the new terms appearing in the two loop counterterms of the 
${\rm 
F}_1$ and ${\rm F}_2$ duals of the simplest 
deformed $SL(3)$ principal model. This speculation is based on a recent 
proposal [15] according to which the naive duality transformation rules for 
the \underbar{renormalized} quantities (as opposed to the \underbar{bare} 
ones) receive perturbative corrections beyond one loop. These corrections are 
encoded into a mapping 
$\gamma_{ij}(g,b)=g_{ij}+b_{ij}+\alpha^\prime M_{ij}(g,b)+\ldots\,$ that 
solves the following equation:
$$
\tilde{T}^{(0)}(g,b)=
\left(\gamma^{-1}\circ T^{(0)}\circ\gamma\right)(\tilde{g},
\tilde{b})\,, 
\equation
$$
where $T^{(0)}(g,b)$ stands for the generalized bare metric, that can be
expressed in terms of the renormalized quantities as:
$$
 T_{ij}^{(0)}(g,b)=g_{ij}+b_{ij}+{\alpha^\prime\over\epsilon}
\hat{R}_{ij}(g,b)+
 {(\alpha^\prime)^2\over\epsilon}\hat{Y}_{ij}(g,b)+...\,,
$$
($\hat{Y}_{ij}={1\over8}Y^{lmk}_{{\phantom{lmk}j}}\hat{R}_{iklm}\,,$)  
and $\tilde{\ }$ denotes the naive dual, given by eq.(3), for both $g,b$ and
\underbar{also for} $T^{(0)}$. 
In ref.[15] eq.(50) was solved for the special, block 
diagonal metric case, when in the adapted coordinate system $b_{ij}\equiv 0$ 
$g_{0\alpha}\equiv 0$ and $g_{00}=g_{00}(\xi^\alpha)$. The examples of the 
${\rm F}_1$ and ${\rm F}_2$ duals we consider here belong to another special 
case when $g_{00}$ is constant but $g_{0\alpha}\ne 0$. Thus we conjecture 
that the new terms that appear are the manifestations of the presence of the 
$\gamma$ mapping, whose general form is not known yet. We hope to return to 
the discussion of this possibility elsewhere.

\noindent{\bf Acknowledgements} This research was supported in part by the 
Hungarian National Science and Research Fund (OTKA) T016251. 
     
\centerline{\bf Appendix A}

In this appendix we describe the explicit form of the two loop $\beta$ 
functions for the various parameters introduced in eq.(42). For the metric
model, eq.(37), they take the form:
$$
\eqalign{\scriptstyle 
\beta_\lambda &={\scriptstyle -{\lambda^{2}\over32\,\pi^{2}}
 (-12\,\pi\,C+24\,\pi-4\,\pi\,BD-16\,\pi\,D-4\,\pi\,B
 )}\cr &-{\scriptstyle {\lambda^{3}\over32\,\pi^{2}} 
(5\,B^{2}+3\,BDC+18\,C^{2}-
9\,D-2\,B+9+3\,DB+21\,DC}\cr &
{\scriptstyle +11\,D^{2}+5\,BD^{2}+3\,BC-9\,C+4\,B^{2}D+2\,B
^{2}D^{2} ),}}
\eqno(A.1)
$$

$$\eqalign{\scriptstyle 
\beta_B &={\scriptstyle B (1+B ) ({\lambda\over64\, (1+D )^{2}\pi^{2}}
 (48\,\pi\,D^{2}+112\,\pi\,D+16
\,\pi\,D^{3}+80\,\pi )}\cr &{\scriptstyle +
{\lambda^{2}\over64\,
 (1+D )^{2}\pi^{2}}
 (-20\,DB-43\,B+4\,BD^{3}+BD^{4}-6\,BD^{2}+52+30\,D+9\,DC}\cr &{\scriptstyle +
6\,D^{2}+3\,C+6\,D^{4}+3\,D^{3}C+2\,D^{3}+9\,D^{2}C ) ),}}
\eqno(A.2)
$$

$$\eqalign{\scriptstyle 
\beta_C &={\scriptstyle 
-{\lambda\over192\,\pi^{2}}
 (-192\,\pi\,C-216\,\pi\,C^{2}-16\,\pi\,BD-128\,\pi\,D-
48\,\pi\,B-192\,\pi\,DC-24\,B\pi\,C-24\,BDC\pi )}
\cr &{\scriptstyle  (-{1\over192\,\pi^{2}}
(-8-104\,D-6\,DB-90\,DC+22\,BD^{2}-18\,BC+28\,B^{2}D
+8\,B^{2}D^{2}+9\,C^{2}DB}\cr &
{\scriptstyle -114\,C-26\,B+32\,B^{2}-90\,C^{2}+16\,D^{2}+
24\,BD^{2}C+24\,B^{2}DC+6\,BDC+12\,B^{2}D^{2}C+4\,D^{3}}\cr &
{\scriptstyle +81\,C^{3}+2\,B
D^{3}+42\,D^{2}C+9\,C^{2}B+72\,C^{2}D+30\,B^{2}C)}\cr &-{\scriptstyle {
1+5\,B^{2}-2\,B\over24\,\pi^{2} (1+D )} )\lambda^{2},}}
\eqno(A.3)
$$

$$\eqalign{\scriptstyle 
\beta_D &={\scriptstyle 
-{\lambda\over64\,\pi^{2}}
 (-24\,\pi\,DC-48\,\pi\,D^{2}-16\,\pi\,BD-16\,\pi\,D-24
\,\pi\,C-8\,\pi\,BD^{2}+24\,\pi\,B )}\cr &{\scriptstyle  (-
{1\over64\,\pi^{2}}(
-2\,B^{2}+18\,DC+6\,BC+6\,B^{2}D+12\,B^{2}D^{2}+7\,BD^{2}}\cr &
{\scriptstyle +36\,C
^{2}-21\,C+12\,BDC+5\,B+8-4\,D+5\,DB+6\,BD^{2}C+39\,D^{2}C}\cr &
{\scriptstyle +7\,BD^{3}+4
\,B^{2}D^{3}+14\,D^{3}+36\,C^{2}D+2\,D^{2})}\cr &+{\scriptstyle {1+5
\,B^{2}-2\,B\over8\,\pi^{2} (1+D )} )\lambda^{2}.}}
\eqno(A.4)
$$   
For the ${\rm F}_1$ dual the corresponding expressions are
$$\eqalign{\scriptstyle 
\beta_\lambda &={\scriptstyle
-{\lambda^{3}\over32\,\pi
^{2}} 
(8-6\,B+18\,C^{2}+D^{2}B^{2}+3\,D^{2}B+2\,DB^{2}+21\,CD
+3\,BDC}\cr &{\scriptstyle 
+2\,B^{2}+10\,D^{2}+3\,BC-5\,D-9\,C+5\,DB )}\cr &{\scriptstyle 
-{\lambda^{2}\over16\,\pi^{2}}
 (12\,\pi-2\,\pi\,B-8\,\pi\,D-2\,\pi\,BD-6\,\pi\,C
 )},}
\eqno(A.5)
$$

$$\eqalign{\scriptstyle 
\beta_B &={\scriptstyle 
-{\lambda^{2}
\over64\,\pi^{2} (1+D )^{2}} 
(1+B ) (12\,D^{3}B^{2}-2\,BD^{4}+22\,D^{2}B^
{2}+20\,DB^{2}+55\,B^{2}-3\,BC+24\,B}\cr &
{\scriptstyle +2\,DB+26\,D^{2}B+14\,D^{3}B-3\,CB
D^{3}-9\,D^{2}BC-9\,BDC+3\,D^{4}B^{2}+64+16\,D^{2}+32\,D )}
\cr &{\scriptstyle -{\lambda\over32\,
\pi^{2} (1+D )^{2}} 
(1+B ) (
-8\,\pi\,D^{3}B-56\,\pi\,BD-24\,\pi\,D^{2}B-40\,\pi\,B )},}
\eqno(A.6)
$$

$$\eqalign{\scriptstyle 
\beta_C &={\scriptstyle 
-{\lambda^{2}\over192\,\pi^{2} (1+D )} 
(60+12\,D-54\,C^{2}+36\,B^{2}+4\,D^{3}B^{2}-18\,B-48\,C
+4\,D^{2}-18\,CD +16\,D^{2}B^{2}}\cr &{\scriptstyle +88\,D^{2}B+
24\,DB^{2}+30\,BC+144\,BDC+
56\,DB+90\,BC^{2}D+2\,BD^{4}+32\,D^{3}B+36\,CBD^{3}}\cr &{\scriptstyle 
+150\,D^{2}BC+4\,D^
{4}+32\,D^{3}+81\,C^{3}+45\,BD^{2}C^{2}+6\,CD^{3}B^{2}+18\,CD^{2}B^{2}}
\cr &{\scriptstyle 
+24\,CDB^{2}+108\,D^{2}C^{2}+90\,D^{2}C+60\,CD^{3}+54\,C^{2}D+12\,CB^{
2}+45\,BC^{2}+81\,DC^{3} )}\cr &{\scriptstyle 
-{\lambda
\over96
\,\pi^{2} (1+D )} 
(-12\,B\pi\,C-24\,\pi\,B-64\,\pi\,D-96\,\pi\,C-24\,B\pi
\,CD-12\,BD^{2}\pi\,C}
\cr &{\scriptstyle -192\,\pi\,CD-64\,\pi\,D^{2}-108\,\pi\,C^{2}-8\,
\pi\,D^{2}B-32\,\pi\,BD-96\,D^{2}\pi\,C-108\,D\pi\,C^{2} )},}
\eqno(A.7)
$$

$$\eqalign{\scriptstyle 
\beta_D &={\scriptstyle 
-{\lambda^{2}
\over64\,\pi^{2} (1+D )} 
(-30-8\,D+36\,C^{2}-24\,B^{2}+8\,D^{3}B^{2}+9\,B-21\,C+
18\,D^{2}-3\,CD}\cr &{\scriptstyle 
+10\,D^{2}B^{2}+24\,D^{2}B+4\,DB^{2}+6\,BC+18\,BDC-2\,D
B+2\,D^{4}B^{2}+7\,BD^{4}+26\,D^{3}B}\cr &{\scriptstyle 
+6\,CBD^{3}+18\,D^{2}BC+16\,D^{4}+
36\,D^{3}+36\,D^{2}C^{2}+57\,D^{2}C+39\,CD^{3}+72\,C^{2}D )}
\cr &{\scriptstyle -{\lambda\over32\,\pi^{2} (
1+D )} 
(4\,\pi\,BD+12\,\pi\,B
-8\,\pi\,D-12\,\pi\,C-12\,D^{2}\pi\,C}\cr &{\scriptstyle 
-24\,\pi\,CD-32\,\pi\,D^{2}-12\,
\pi\,D^{2}B-24\,\pi\,D^{3}-4\,\pi\,D^{3}B )},}
\eqno(A.8)
$$
while for the ${\rm F}_2$ dual they are:
$$\eqalign{\scriptstyle 
\beta_\lambda &={\scriptstyle 
 (- (10+2\,B^{2}+5\,B )D^{2}-
 (4\,B^{2}+3\,CB-9+15\,C+3\,B )D
 }\cr &{\scriptstyle -
(18-2\,B-9\,C+5\,B^{2}+3\,CB+9\,C^{2} )){\lambda^{3
}\over32\,\pi^{2}}}\cr &{\scriptstyle 
+ ( (2\,\pi\,B+8\,\pi )D-(
12\,\pi-6\,\pi\,C-2\,\pi\,B) ){\lambda^{2}\over16\,\pi^{2}}},}
\eqno(A.9)
$$

$$\eqalign{\scriptstyle 
\beta_B &={\scriptstyle 
 ({B\over64\,\pi^{2}} 
 (2\,D^{2}+3\,BD^{2}-9\,CD+B^{2}D^{2}+2\,B^{2}D-
26\,D-9\,CBD-24\,BD-9\,C-3\,B-9\,CB+8-11\,B^{2} )}\cr &{\scriptstyle 
+{B (1-B^{2} )\over2\pi^{2} (D+1 )^{2}}
 )\lambda^{2}
+ ({B (8\,\pi\,BD+8\,\pi+8\,\pi\,D+8\,
\pi\,B )\over32\,\pi^{2}}+{B (1+B )\over\pi\, 
(D+1 )} )\lambda},}
\eqno(A.10)
$$

$$\eqalign{\scriptstyle 
\beta_C &={\scriptstyle 
 (-{1\over192\,\pi^{2}} 
 (124-24\,D-26\,B+252\,C+32\,B^{2}+198\,C^{2}-18\,CB+28
\,D^{2}+30\,CB^{2}}\cr &{\scriptstyle 
+4\,D^{3}+135\,C^{3}+60\,CD^{2}+144\,C^{2}D+2\,BD^{3
}+9\,C^{2}B+24\,CBD^{2}+9\,C^{2}BD+12\,CB^{2}D^{2}}\cr &{\scriptstyle 
+24\,CB^{2}D+22\,BD^
{2}+8\,B^{2}D^{2}+78\,CD+6\,CBD-6\,BD+28\,B^{2}D)}\cr &{\scriptstyle 
-{
5\,B^{2}-2\,B+1\over24\,\pi^{2} (D+1 )} )\lambda^{2}}
\cr &{\scriptstyle 
-{\lambda\over96\,\pi^{2}}
 (-24\,\pi\,B-12\,C\pi\,B-96\,\pi\,C-8\,\pi\,BD-12\,C\pi\,
BD-96\,\pi\,CD-108\,\pi\,C^{2}-64\,\pi\,D )},}
\eqno(A.11)
$$

$$\eqalign{\scriptstyle 
\beta_D &={\scriptstyle 
 (-{1\over64\,\pi^{2}}
 (-2\,B^{2}+18\,C^{2}+42\,D+6\,CB+38+7\,BD^{2}+6\,B^{2}D
+12\,B^{2}D^{2}+5\,B+30\,CD-9\,C+12\,CBD}\cr &{\scriptstyle 
+16\,D^{3}+20\,D^{2}+39\,CD^{2
}+18\,C^{2}D+7\,BD^{3}+6\,CBD^{2}+4\,B^{2}D^{3}+5\,BD)
+{
5\,B^{2}-2\,B+1\over8\,\pi^{2} (D+1 )} )\lambda^{2}
}\cr &{\scriptstyle -{\lambda\over32\,\pi^{2}}
 (-4\,\pi\,BD^{2}-12\,\pi\,C-8\,\pi\,BD+12\,\pi\,B-12\,\pi
\,CD-24\,\pi\,D^{2}-8\,\pi\,D )}.}
\eqno(A.12)
$$

\centerline{\bf Appendix B}

In this appendix we show on the example of the perturbative subspace of the
${\rm F}_1$/${\rm F}_2$ duals 
$$
\hat{b}=\hat{c}+{7\over3}\hat{d}-{\lambda\over\pi}(2\hat{c}+
{41\over4}\hat{d}),
\eqno(B.1)
$$
that the two loop corrections to $N_i$ are indeed of higher order in $\lambda$. 
On the surface defined by (B.1) in the space of $\hat{b},\hat{c},\hat{d}$ 
the appropriately linearized renormalization equations have the following 
form:
$$\eqalign{ 
{d\lambda\over dt}&=-{\lambda^2\over8\pi}(\alpha_0c+\beta_0d+L)
-{\lambda^3\over8\pi^2}(\alpha_1c+\beta_1d+S),\cr
{dc\over dt}&={\lambda\over8\pi}\Delta_1c-{\lambda^2\over8\pi^2}
\tilde{\Delta}_1c,\cr
{dd\over dt}&={\lambda\over8\pi}\Delta_2d-{\lambda^2\over8\pi^2}
\tilde{\Delta}_2d,}
\eqno(B.2)
$$
where all terms of $o(c^2,cd,d^2)$ are neglected and the various constants 
(whose actual values are not important for the rest) can be obtained from 
eq.(46) and (A.5-A.12). One verifies by direct computation that the 
expressions:
$$\eqalign{
\tilde{N}_1={\lambda\over c^{-L/\Delta_1}}\exp\bigl({\alpha_0\over\Delta_1}c
+{\beta_0\over\Delta_2}d\bigr)\bigl(1+\lambda (P+k_1c+k_2d)\bigr),\cr
\tilde{N}_2={\lambda\over d^{-L/\Delta_2}}\exp\bigl({\alpha_0\over\Delta_1}c
+{\beta_0\over\Delta_2}d\bigr)\bigl(1+\lambda (Q+l_1c+l_2d)\bigr),}
\eqno(B.3)
$$
where
$$\eqalign{
P&=-{1\over\pi}(S+L{\tilde{\Delta}_1\over\Delta_1}),\quad
k_1={1\over\pi(\Delta_1-L)}\bigl( \alpha_0\pi P
+\alpha_1+\alpha_0{\tilde{\Delta}_1\over\Delta_1}\bigr),\cr
k_2&={1\over\pi(\Delta_2-L)}\bigl( \beta_0\pi P
+\beta_1+\beta_0{\tilde{\Delta}_2\over\Delta_2}\bigr),}
$$
and
$$\eqalign{
Q&=-{1\over\pi}(S+L{\tilde{\Delta}_2\over\Delta_2}),\quad
k_1={1\over\pi(\Delta_1-L)}\bigl( \alpha_0\pi Q
+\alpha_1+\alpha_0{\tilde{\Delta}_1\over\Delta_1}\bigr),\cr
k_2&={1\over\pi(\Delta_2-L)}\bigl( \beta_0\pi Q
+\beta_1+\beta_0{\tilde{\Delta}_2\over\Delta_2}\bigr),}
$$
are indeed invariants of the flow described by (B.2), provided we neglect 
$c^2$, $cd$, and $d^2$.  
From (B.3) one obtaines $p_i\sim {\lambda\over 
\tilde{N}_i}+o(\lambda^2)$, with
$p_1=\hat{c}$, $p_2=\hat{d}$. The $
\tilde{N}_i$ and the $\lambda\rightarrow 0$ 
behaviour of the parameters in the perturbative 
subspace of the metric model are derived in an entirely analogous way.

\centerline{\bf References}

\item{[1]} T.~Buscher, {\sl Phys. Lett.}
{\bf B194} (1987) 51.

\item{[2]} 
E.~Alvarez, L.~Alvarez-Gaum\'e and Y.~Lozano, {\sl Nucl.~Phys.~B}
(Proc.~Suppl.) {\bf 41} (1995) 1; (hep-th/9410237).

\item{[3]} X.~De la Ossa and E.~Quevedo, {\sl Nucl. Phys.}
{\bf B403} (1993) 377.
\item{} 
E.~Alvarez, L.~Alvarez-Gaum\'e and Y.~Lozano, {\sl Nucl. Phys.}
{\bf B424} (1994) 155.

\item{[4]} E.~Alvarez, L.~Alvarez-Gaum\'e and Y. ~Lozano, {\sl Phys. Lett.}
{\bf B336} (1994) 183.

\item{[5]} T.~Curtright and C.~Zachos, {\sl Phys. Rev.}
{\bf D49} (1994) 5408, and hep-th/9407044. 
Y.~Lozano  {\sl Phys. Lett.} {\bf B355} (1995) 165.

\item{[6]} E.~Fradkin and A.A.~Tseytlin, {\sl Ann. Phys.}
{\bf 162} (1985) 31.

\item{[7]} J.~Balog, P.~Forg\'acs, Z.~Horv\'ath, L.~Palla, hep-th/9601091
{\sl Nucl. Phys. B} (Proc. Suppl.){\bf 49} (1996) 16.

\item{[8]} P.E. Haagensen, hep-th/9604136

\item{[9]} C.M.~Hull and P.K.~Townsend, {\sl Phys. Lett.}
{\bf B191} (1987) 115.

\item{[10]} R.R.~Metsaev and A.A.~Tseytlin, {\sl Phys. Lett.}
{\bf B191} (1987) 354.

\item{[11]} H.~Osborn, {\sl Ann. Phys.}
{\bf 200} (1990) 1.

\item{[12]} P.S.~Howe, G.~Papadopoulos and K.S.~Stelle, {\sl Nucl. Phys.}
{\bf B296} (1988) 26.
\item{} H.~Osborn, {\sl Nucl. Phys.}
{\bf B294} (1987) 595.

\item{[13]} J.~Balog, P.~Forg\'acs, Z.~Horv\'ath, L.~Palla, unpublished.

\item{[14]} D.~Friedan, {\sl Phys. Rev. Lett} {\bf 45} (1980) 1057,
{\sl Ann. Phys.} {\bf 163} (1985) 318.

\item{[15]} J.~Balog, P.~Forg\'acs, Z.~Horv\'ath, L.~Palla, hep-th/9606187, 
to appear in {\sl Phys. Lett.} {\bf B}.

\vfill
\bye